\documentclass[twocolumn,english,pra,aps,superscriptaddress,showpacs]{revtex4}
\usepackage[T1]{fontenc}
\usepackage[latin9]{inputenc}
\usepackage{amsmath}
\usepackage{amssymb}
\usepackage{graphicx}
\usepackage{esint}

\makeatletter
\@ifundefined{textcolor}{}
{%
 \definecolor{BLACK}{gray}{0}
 \definecolor{WHITE}{gray}{1}
 \definecolor{RED}{rgb}{1,0,0}
 \definecolor{GREEN}{rgb}{0,1,0}
 \definecolor{BLUE}{rgb}{0,0,1}
 \definecolor{CYAN}{cmyk}{1,0,0,0}
 \definecolor{MAGENTA}{cmyk}{0,1,0,0}
 \definecolor{YELLOW}{cmyk}{0,0,1,0}
}

\makeatother

\usepackage{babel}
\begin{document}

\title{Quantum information with conserved quantities}

\author{Jing Liu}
\affiliation{Zhejiang Institute of Modern Physics, Department of Physics, Zhejiang
University, Hangzhou 310027, China}

\author{Jing Cheng}
\affiliation{Department of Physics, South China University of Technology, Guangzhou 510640, China}

\author{Li-Bin Fu}
\affiliation{Institute of Applied Physics and Computational Mathematics, Beijing
100088, China}

\author{Xiaoguang Wang}
\email{xgwang@zimp.zju.edu.cn}
\affiliation{Zhejiang Institute of Modern Physics, Department of Physics, Zhejiang
University, Hangzhou 310027, China}
\affiliation{Synergetic Innovation Center of Quantum Information and Quantum Physics, University of Science and Technology of China, Hefei, Anhui 230026,
China}

\begin{abstract}
Conserved quantities are crucial in quantum physics. Here we discuss a general
scenario of Hamiltonians. All the Hamiltonians within this scenario
share a common conserved quantity form. For unitary parametrization
processes, the characteristic operator of this scenario is analytically
provided, as well as the corresponding quantum Fisher information
(QFI). As the application of this scenario, we focus on two classes
of Hamiltonians: $\mathfrak{su}(2)$ category and canonical category.
Several specific physical systems in these two categories are discussed
in detail. Besides, we also calculate an alternative form of QFI
in this scenario.
\end{abstract}


\maketitle

\section{Introduction}

With the development of quantum physics, quantum mechanics nowadays
is way beyond the phase of logical argument and mind experiments,
but deep into the applied field and even our daily lives. Quantum
metrology is one of the most successful applications of quantum mechanics.
Starting from the pioneer work of Caves~\cite{Caves,Braunstein},
people gradually realized that many quantum effects and states~\cite{Giovannetti,Genoni,Pezze,Chin,Gerry,Anisimov,Braun,Joo,
Gerry1,Erol,Jarzyna,XMLu,Toth}
are available to enhance the measure precision of physical parameters
via various quantum systems.

Quantum Fisher information (QFI) is a central quantity in quantum
metrology because it depicts the lower bound on the variance of the
estimator $\hat{\theta}$ due to the Cram\'{e}r-Rao theory: $\delta^{2}\hat{\theta}\geq(\nu F)^{-1}$~\cite{Helstrom,Holevo},
where $\delta^{2}\hat{\theta}$ is the variance, $\nu$ is the number
of repeated experiments, and $F$ is the QFI. Generally, the QFI is
defined as $F=\mathrm{Tr}(\rho L^{2})$. Here $\rho$ is a parametrized
state and $L$ is the symmetric logarithmic derivative (SLD), which
is determined by the equation $\partial_{\theta}\rho=(\rho L+L\rho)/2$.
$\theta$ is the parameter under estimation. Quantum Fisher information
matrix (QFIM) $\mathcal{F}$ is the counterpart of QFI in multi-parameter
estimation. The element of QFIM is defined as $\mathcal{F}_{mn}=\mathrm{Tr}(\rho\{L_{m},L_{n}\})/2$.
$L_{m}$, $L_{n}$ are the SLD operators for $m$th and $n$th parameters
under estimation, respectively.

Because of the importance of QFI, its calculation is always an interesting
and attractive topic. Recently, several alternative formulas of QFI
have been developed~\cite{Pang,JLiu_H,ZJiang,JLiu1,JLiu2,GRJin,JLiu3}.
For unitary parametrization processes, new expressions of QFI and QFIM were
given for both pure and mixed states~\cite{Pang,JLiu_H}. The QFI and QFIM 
here are totally determined by a characteristic operator $\mathcal{H}$
and the initial states. In these expressions, all the information
of parametrization is involved in $\mathcal{H}$. Furthermore, $\mathcal{H}$
can be written into an expanded form~\cite{JLiu_H}, which is particularly
useful when the $n$th order commutation between the Hamiltonian and
its partial derivative can be truncated or is periodic. Moreover,
for a general parametrized exponential state, the SLD operator is
also found to be expressed in an expanded form~\cite{ZJiang}. These
facts prompt us to study various Hamiltonians owning above properties.
This is the major motivation of this paper.

Conserved quantities are crucial in both classical and quantum physics.
Since Emmy Noether connected them with differentiable symmetries in
Noether's theorem, searching for conserved quantities becomes a prior
mission when facing a novel new system. Locating conserved quantities
will not only help us to find hidden symmetries of these systems,
but also give us a easy perspective to describe and classify them.

In this paper, we discuss a scenario, in which all systems share a
common conserved quantity form. For the unitary parametrization processes,
we provide the analytical expressions of the characteristic operator
and the QFI. The maximum QFI and the corresponding optimal initial
states are discussed. Moreover, we also study QFI for the parametrized
thermal states in these systems. The scenario we discuss includes
many systems, of which two typical classes, $\mathfrak{su}(2)$ category
and canonical category, are the main applications of this paper. In
the $\mathfrak{su}(2)$ category, ferromagnetic two-spin system, anisotropic
two-spin system and a spin-one system are discussed in detail. In
the canonical category, a cavity optomechanical system are provided
as an example. At the end of this paper, we discuss an alternative
form of QFI given by Luo \emph{et al.}~\cite{SLuo}. The formula
of this alternative QFI for above scenario is analytically calculated
and discussed.

The paper is organized as follows. In Sec.~\ref{sec:Theory}, we
review quantum metrology with unitary parametrization processes. In
Sec.~\ref{sec:conserved-quantities}, we propose a scenario in which
all systems share a common conserved quantity form. For unitary parametrization
processes, the characteristic operator and the QFI are analytically
provided. We also discussed the QFI for the parametrized thermal states
in the scenario. In Sec.~\ref{sec:Applications}, the applications
of this scenario, including two typical classes: $\mathfrak{su}(2)$
category and canonical category, are given and discussed. In Sec.~\ref{sec:Alternative}, an alternative form of QFI is calculated 
for this scenario. Section~\ref{sec:Conclusion}
is the conclusion of this paper.

\section{Unitary parametrization\label{sec:Theory}}

The unitary parametrization process is a widely used parametrization
strategy in quantum metrology. Recently, it has been found that the
QFI and QFIM for both pure and mixed states can be expressed via a
characteristic function $\mathcal{H}$~\cite{Pang,JLiu_H}. Denoting
the parameter under estimation as $\theta$, $\mathcal{H}$ is defined
as~\cite{Boixo,Taddei,Pang,JLiu_H}
\begin{equation}
\mathcal{H}_{\theta}:=i\left(\partial_{\theta}U^{\dagger}\right)U,
\end{equation}
where $U$ is a unitary parametrization transformation, i.e., $\rho_{\theta}=U\rho_{0}U^{\dagger}$
with $\rho_{0}$ a $\theta$-independent density matrix. Now we consider
the situation that the transformation is generated by a time-independent
Hamiltonian, namely, $U$ can be written in the form
\begin{equation}
U=\exp[-itH(\vec{\theta})],
\end{equation}
where $\vec{\theta}=(\theta_{1},\theta_{2},...)^{\mathrm{T}}$ is
a vector of parameters under estimation and the parametrized Hamiltonian
$H(\vec{\theta})$ is time-independent. Based on a recent work~\cite{JLiu_H},
the characteristic operator $\mathcal{H}_{\theta}$ for parameter
$\theta$ can be expressed in an expanded form
\begin{equation}
\mathcal{H}_{\theta}=i\sum_{n=0}^{\infty}\frac{(it)^{n+1}}{(n+1)!}
\left(H^{\times}\right)^{n}\partial_{\theta}H,\label{eq:H}
\end{equation}
where $H^{\times}=[H,\cdot]$ is a superoperator. With a known $\mathcal{H}$,
the QFIM can be obtained easily. The element of QFIM can be expressed
by~\cite{JLiu_H}
\begin{eqnarray}
\mathcal{F}_{mn} & \!\!=\!\! & \sum_{i=1}^{M}4p_{i}\mathrm{cov}_{i}\left(\mathcal{H}_{m},\mathcal{H}_{n}\right)
\nonumber \\
&  & -\sum_{i\neq j}\frac{8p_{i}p_{j}}{p_{i}+p_{j}}\mathrm{Re}\left(\langle\psi_{i}
|\mathcal{H}_{m}|\psi_{j}\rangle\langle\psi_{j}|\mathcal{H}_{n}
|\psi_{i}\rangle\right)\!,\label{eq:F_mn}
\end{eqnarray}
where the covariance reads
\begin{equation}
\mathrm{cov}_{i}\left(\mathcal{H}_{m},\mathcal{H}_{n}\right)
=\frac{1}{2}\langle\{\mathcal{H}_{m},\mathcal{H}_{n}\}\rangle_{i}
-\langle\mathcal{H}_{m}\rangle_{i}\langle\mathcal{H}_{n}\rangle_{i}.
\end{equation}
In above equations, $\mathcal{H}_{m}$ is short for $\mathcal{H}_{\theta_{m}}$.
$p_{i}$ and $|\psi_{i}\rangle$ are the $i$th eigenvalue and eigenstate
of initial state $\rho_{0}$, which means $p_{i}$ and $|\psi_{i}\rangle$
are independent of $\theta$. $M$ is the dimension of the support
of $\rho_{0}$. $\langle\cdot\rangle_{i}$ is the expected value on
$|\psi_{i}\rangle$, i.e., $\langle\cdot\rangle_{i}=\langle\psi_{i}|\cdot|\psi_{i}\rangle$.
It is known that the diagonal elements of QFIM are the QFIs for corresponding
single-parameter estimations, thus, the QFI for a unitary parametrization
process can be written into the form~\cite{JLiu_H}
\begin{equation}
F_{\theta}=\sum_{i=1}^{M}4p_{i}\langle\Delta^{2}
\mathcal{H}_{\theta}\rangle_{i}-\sum_{i\neq j}\frac{8p_{i}p_{j}}{p_{i}+p_{j}}|\langle\psi_{i}
|\mathcal{H}_{\theta}|\psi_{j}\rangle|^{2},\label{eq:F_mixed}
\end{equation}
where $\langle\Delta^{2}\mathcal{H}_{\theta}\rangle_{i}:=\langle
\mathcal{H}_{\theta}^{2}\rangle_{i}-\langle\mathcal{H}_{\theta}\rangle_{i}^{2}$
is the variance of the characteristic operator on $|\psi_{i}\rangle$.
For a purely initial state, the element of QFIM reduces to~\cite{JLiu_H}
\begin{equation}
\mathcal{F}_{mn}=4\mathrm{cov}\left(\mathcal{H}_{m},\mathcal{H}_{n}\right).
\label{eq:QFIM_pure}
\end{equation}
The covariance is taken on the initial state. Based on this equation,
the QFI for purely initial states is ~\cite{Pang,JLiu_H}
\begin{equation}
F_{\theta}=4\langle\Delta^{2}\mathcal{H}_{\theta}\rangle,
\label{eq:F_pure}
\end{equation}
namely, the QFI for parameter $\theta$ is actually the variance of
the corresponding characteristic operator.

In the expression of QFIM in Eq.~(\ref{eq:F_mn}), for any $m$,
if we assume $\mathcal{H}_{m}=\mathcal{H}_{m}^{\prime}+c_{m}$, with
$c_{m}$ a complex number, it can be checked that $\mathrm{cov}_{i}(\mathcal{H}_{m},\mathcal{H}_{n})
=\mathrm{cov}_{i}(\mathcal{H}_{m}^{\prime},\mathcal{H}_{n}^{\prime})$.
Meanwhile, in the second part of Eq.~(\ref{eq:F_mn}), the overlap
$\langle\psi_{i}|c_{m}|\psi_{j}\rangle$ always vanishes for any $m$
when $i\neq j$. Thus, $\mathcal{H}_{m}$ and $\mathcal{H}_{m}^{\prime}$
share the same expression of QFIM. This fact indicates that the number
terms of characteristic operator do not affect the value of QFIM,
therefore they can be neglected in metrological problems.

\section{conserved quantities\label{sec:conserved-quantities}}

Conserved quantities are important in theoretical physics.
In the following we propose a scenario in which all systems share
a common conserved quantity. At first, we introduce a $\theta$-dependent
Hermitian operator $\mathcal{V}$, which is defined as
\begin{equation}
\mathcal{V}=\left[\left(H^{\times}\right)^{2}-\Omega^{2}\right]
\partial_{\theta}H,\label{eq:V_definition}
\end{equation}
where $\Omega^{2}$ is a real number and $\theta$ is a parameter
in the Hamiltonian. This operator is generated by the Hamiltonian
$H$ and related to the parameter $\theta$. The central quality of
the scenario we discuss here is that $\mathcal{V}$ is a conserved quantity
for all systems in this scenario, namely, the Hamiltonian $H$ satisfies
\begin{equation}
\left[\mathcal{V},H\right]=0.
\end{equation}
Based on the definition of $\mathcal{V}$, above equation can be rewritten
into
\begin{equation}
\left[\left(H^{\times}\right)^{2}-\Omega^{2}\right]H^{\times}
\partial_{\theta}H=0.\label{eq:assume}
\end{equation}
This equation implies that $H^{\times}\partial_{\theta}H$ is the
eigenoperator of superoperator $(H^{\times})^{2}$, with $\Omega^{2}$
the corresponding eigenvalue. From this aspect, one can check that
$H^{\times}\partial_{\theta}H$ and $(H^{\times})^{2}\partial_{\theta}H$
are also eigenoperators of $(H^{\times})^{2n}$ with $\Omega^{2n}$
the eigenvalues for $n\geq0$, i.e.,
\begin{equation}
\left[(H^{\times})^{2n}-\Omega^{2n}\right](H^{\times})^{i}\partial_{\theta}H=0,
\end{equation}
where $i=1,2$. In the following we take $\theta$ as the parameter
under estimation. Based on this equation, the characteristic operator
$\mathcal{H}_{\theta}$ in Eq.~(\ref{eq:H}) will be separated into
two parts via the parity of $n$. Through some straightforward calculations,
the analytical expression of $\mathcal{H}_{\theta}$ can be obtained
as below
\begin{eqnarray}
\mathcal{H}_{\theta} & = & \left[-t-i(\partial_{t}f)H^{\times}+f(H^{\times})^{2}\right]
\partial_{\theta}H.\label{eq:H_theta}
\end{eqnarray}
Here $f$ is short for the function $f(\Omega,t)$, which is defined as
\begin{equation}
f(\Omega,t):=\frac{1}{\Omega^{3}}\left[\Omega t
-\sin\left(\Omega t\right)\right].
\end{equation}
If $\Omega>0$, $f$ is positive and monotone increasing with the
passage of time. Utilizing the conserved quantity $\mathcal{V}$,
the expression of $\mathcal{H}_{\theta}$ can be rewritten into
\begin{equation}
\mathcal{H}_{\theta}=f\mathcal{V}+\left(f\Omega^{2}-t\right)
\partial_{\theta}H-i\left(\partial_{t}f\right)H^{\times}\partial_{\theta}H.
\end{equation}
Moreover, from the expression of $f$, the coefficients can be simplified
as $f\Omega^{2}-t=-\Omega^{-1}\sin\left(\Omega t\right)$ and $\partial_{t}f=2\Omega^{-2}\sin^{2}\left(\Omega t/2\right)$,
then $\mathcal{H}_{\theta}$ can be finally expressed by
\begin{equation}
\mathcal{H}_{\theta}=f\mathcal{V}\!-\frac{1}{\Omega}\sin\!\left(\Omega t\right)\!\partial_{\theta}H-\frac{i2}{\Omega^{2}}\sin^{2}
\!\left(\frac{\Omega}{2}t\right)\! H^{\times}\partial_{\theta}H.
\label{eq:H_temp}
\end{equation}

This is the general expression of the characteristic operator $\mathcal{H}_{\theta}$
for the scenario in which $\mathcal{V}$ is a conserved quantity.
In this scenario, the characteristic operator is the linear combination
of the conserved quantity $\mathcal{V}$, partial derivative $\partial_{\theta}H$ and commutation between $\partial_{\theta}H$
and $H$. With the expression of $\mathcal{H}_{\theta}$, the QFIM and QFI
can be calculated through Eqs.~(\ref{eq:F_mn})-(\ref{eq:F_pure})
for both mixed and pure states. For the long-time limit or
the situations that the value of $\Omega$ is very large, the oscillating
terms in Eq.~(\ref{eq:H_temp}) can be neglected and $\mathcal{H}_{\theta}$
reduces to
\begin{equation}
\mathcal{H}_{\theta}=\frac{t}{\Omega^{2}}\mathcal{V}.
\end{equation}
The characteristic operator is then proportional to $\mathcal{V}$.
For purely initial states, the QFI is
\begin{equation}
F_{\theta}=\frac{t^{2}}{\Omega^{4}}\langle\Delta^{2}\mathcal{V}\rangle,
\end{equation}
namely, it is actually determined by the fluctuation of the conserved
quantity.

When $\mathcal{V}$ is a number (which is a trivial conserved quantity),
it can be neglected according
to the analysis in Sec.~\ref{sec:Theory}. Thus, the characteristic
operator in this case reduces to
\begin{equation}
\mathcal{H}_{\theta}=-\frac{1}{\Omega}\!\left[\sin\left(\Omega t\right)\partial_{\theta}H-\frac{i2}{\Omega}\sin^{2}
\!\left(\frac{\Omega}{2}t\right)\! H^{\times}\partial_{\theta}H\right].
\end{equation}
For purely initial states, the corresponding QFI is
\begin{eqnarray}
F_{\theta} & \!\!=\!\! & \frac{1}{\Omega^{2}}\!\Big[\sin^{2}\!\left(\Omega t\right)\!\big\langle(\partial_{\theta}H)^{2}\big\rangle
-\frac{4}{\Omega^{2}}\sin^{4}\!\!\left(\!\frac{\Omega}{2}t\!\right)
\!\!\big\langle(H^{\times}\partial_{\theta}H)^{2}\big\rangle\nonumber \\
&  & -\frac{i4}{\Omega}\sin\left(\Omega t\right)\sin^{2}\!\left(\frac{\Omega}{2}t\right)\!\big\langle H^{\times}(\partial_{\theta}H)^{2}\big\rangle\Big],
\end{eqnarray}
where the equality~$\{\partial_{\theta}H,H^{\times}\partial_{\theta}H\}
=H^{\times}(\partial_{\theta}H)^{2}$~has been applied. 
$\{\cdot,\cdot\}$ represents the anticommutation.
A simple example of this scenario is that $H^{\times}\partial_{\theta}H$
is proportional to $\partial_{\theta}H$, i.e., $H^{\times}\partial_{\theta}H=\Omega\partial_{\theta}H.$
In this example, $\mathcal{V}=0$, then $\mathcal{H}$ reduces to
$i\Omega^{-1}(e^{i\Omega t}-1)\partial_{\theta}H$~\cite{JLiu_H}.
Especially, if $\Omega$ can be chosen as $i\omega$ with $\omega$
a positive number in this example, the characteristic  operator~$\mathcal{H}_{\theta}$~will reduce to~$-\partial_{\theta}H/\omega$~for
the long-time limit. A realistic case of this scenario is a collective spin system in an external magnetic field, which has been discussed in Ref.~\cite{JLiu_H}
in detail.

Moreover, if the operator $\mathcal{V}_{1}=(H^{\times}-\Omega)\partial_{\theta}H$
is a non-trivial conserved quantity, then $\mathcal{V}=\Omega\mathcal{V}_{1}$
is also a non-trivial conserved quantity. Therefore the characteristic
operator in this case can also be expressed in the form of Eq.~(\ref{eq:H_temp}).

\emph{Thermal states of the scenario}.-Thermal states widely appear
in realistic world. In quantum theory, a general thermal state can
be written as
\begin{equation}
\rho=\frac{1}{Z}\exp\left(-\beta H\right),\label{eq:thermal}
\end{equation}
where $\beta=1/(k_{\mathrm{B}}T)$ and the partition function $Z=\mathrm{Tr}(e^{-\beta H})$.
$T$ is the temperature and $k_{\mathrm{B}}$ is the Boltzmann constant.
In Plank unit, $k_{\mathrm{B}}=1$. Recently, Jiang~\cite{ZJiang}
provides the expression of SLD for a general exponential state $\rho_{\theta}=\exp[G(\theta)]$,
which is
\begin{equation}
L=\sum_{n=0}^{\infty}\frac{4\left(4^{n+1}-1\right)
\mathcal{B}_{2n+2}}{\left(2n+2\right)!}\left(G^{\times}\right)^{2n}
\partial_{\theta}G.\label{eq:L}
\end{equation}
Here $\mathcal{B}_{2n+2}$ is the $(2n+2)$th Bernoulli number. For
the thermal state expressed in~(\ref{eq:thermal}), $G=-\beta H-\ln Z$.
Then it is easy to check the equality $(G^{\times})^{2n}\partial_{\theta}G=(-\beta)^{2n+1}
(H^{\times})^{2n}\partial_{\theta}H$.
For any system in the scenario, $\mathcal{V}$ is a conserved quantity. 
The SLD operator can then be calculated as
\begin{eqnarray}
L & = & \beta\left[r\mathcal{V}+\left(r\Omega^{2}-1\right)
\partial_{\theta}H\right],
\label{eq:L_thermal}
\end{eqnarray}
where $r$ is short for $r(\beta,\Omega)$, which is defined as
\begin{equation}
r(\beta,\Omega):=\frac{1}{\Omega^{2}}\left[1-\frac{1}{\beta\Omega}\tanh
\left(\beta\Omega\right)\right].\label{eq:R}
\end{equation}
The regime of $r$ is $[0,\Omega^{-2}]$. Meanwhile, one can see that
$r\Omega^{2}-1=-\tanh\left(\beta\Omega\right)/(\beta\Omega).$ Then
the SLD can be alternatively written as
\begin{equation}
L=\beta r\mathcal{V}-\frac{1}{\Omega}\tanh\left(\beta\Omega\right)\partial_{\theta}H.
\end{equation}
With above equation, the QFI $F_{\mathrm{T}}=\langle L^{2}\rangle$
can be finally expressed by
\begin{eqnarray}
F_{\mathrm{T}} & = & \beta^{2}r^{2}\langle\mathcal{V}^{2}\rangle_{\mathrm{T}}
+\frac{1}{\Omega^{2}}\tanh^{2}\left(\beta\Omega\right)\big\langle
\!\left(\partial_{\theta}H\right)^{2}\!\big\rangle_{\mathrm{T}}\nonumber \\
&  & -\frac{\beta}{\Omega}r\tanh\left(\beta\Omega\right)\big\langle\!\left\{ \mathcal{V},\partial_{\theta}H\right\} \!\big\rangle_{\mathrm{T}}.
\end{eqnarray}
Here $\langle\cdot\rangle_{\mathrm{T}}$
is the expected value on the thermal states. If the thermal states
in Eq.~(\ref{eq:thermal}) can be rewritten into the form $Ue^{-\beta H_{0}}U^{\dagger}$,
with $H_{0}$ a parameter independent Hamiltonian, the QFI can also
be calculated utilizing the function $\mathcal{H}_{\theta}$ in Eq.~(\ref{eq:H_temp}).
At the zero-temperature limit, $\tanh(\beta\Omega)=1$ and $r=\Omega^{-2}$,
the SLD reduces to $L=\beta\Omega^{-2}\mathcal{V}$ and the QFI is
$F_{\mathrm{T}}=\beta^{2}\Omega^{-4}\langle\mathcal{V}^{2}\rangle_{\mathrm{T}}$.
Similarly, at the high-temperature limit, $\tanh(\beta\Omega)\simeq\beta\Omega$
and $r=0$, the SLD is $L=-\beta\partial_{\theta}H$ and the QFI can
be written as $F_{\mathrm{T}}=\beta^{2}\langle(\partial_{\theta}H)^{2}\rangle_{\mathrm{T}}$.

\section{Applications\label{sec:Applications}}

In the following we will solve the metrological problems in some realistic
systems in the scenario where $\mathcal{V}$ is a conserved quantity.
We mainly focus on two classes of Hamiltonians. The first one is mainly
related to the generators of $\mathfrak{su}(2)$ algebra and is called
$\mathfrak{su}(2)$ category; the second one is related to the canonical
variables: the position operator $x$ and the momentum operator $p$, and
it is called canonical category. We first discuss $\mathfrak{su}(2)$
category.

\subsection{$\mathfrak{su}(2)$~category}

\subsubsection*{Ferromagnetic two-spin system}

As the first application of the $\mathfrak{su}(2)$ category, we now
consider a ferromagnetic two-spin system in an external magnetic field.
The Hamiltonian of this system reads
\begin{equation}
H_{1}=-\sigma_{1}^{x}\sigma_{2}^{x}-B\left(\sigma_{1}^{z}
+\sigma_{2}^{z}\right),\label{eq:H_1}
\end{equation}
where $\sigma_{1}^{i}=\sigma_{i}\otimes\openone$ and $\sigma_{2}^{i}=\openone\otimes\sigma_{i}$
for $i=x,y,z$. $\sigma_{i}$ is a Pauli matrix, $\openone$ is the
identity matrix, and $B$ is the strength of the external field.
The optimization of QFI in a general Ising model with GHZ-type state 
has been discussed recently~\cite{Skotiniotis}.
In this case, we take $B$ as the parameter under estimation. Before
the main calculation, we introduce three operators
\begin{equation}
\begin{cases}
J_{x}=\!\! & \frac{1}{4}\left(\sigma_{1}^{x}\sigma_{2}^{x}-\sigma_{1}^{y}\sigma_{2}^{y}\right),\\
J_{y}=\!\! & \frac{1}{4}\left(\sigma_{1}^{x}\sigma_{2}^{y}+\sigma_{1}^{y}\sigma_{2}^{x}\right),\\
J_{z}=\!\! & \frac{1}{4}\left(\sigma_{1}^{z}+\sigma_{2}^{z}\right).
\end{cases}
\end{equation}
It is worth to notice that $J_{x}$, $J_{y}$ and $J_{z}$ satisfy
the $\mathfrak{su}(2)$ commutation $\left[J_{i},J_{j}\right]=i\epsilon_{ijk}J_{k}$
with $\epsilon_{ijk}$ the Levi-Civita symbol. In addition, the anticommutation
is $\{J_{i},J_{j}\}=2\delta_{ij}J_{i}$ with $\delta_{ij}$ the Kronecker
delta function. Using these operators, the first and second order
commutations between the Hamiltonian and its derivative are calculated as below
\begin{eqnarray}
H_{1}^{\times}\partial_{B}H_{1} & = & -i8J_{y},\\
\left(H_{1}^{\times}\right)^{2}\partial_{B}H_{1} & = & 16\left(2BJ_{x}-J_{z}\right).
\end{eqnarray}
Utilizing above two equations, we can obtain
\begin{equation}
\left[\left(H_{1}^{\times}\right)^{2}-4\left(1+4B^{2}\right)^{2}\right]
H_{1}^{\times}\partial_{\theta}H_{1}=0.
\end{equation}
This equation implies that if we choose
\begin{equation}
\Omega=2\sqrt{1+4B^{2}},
\end{equation}
the operator $\mathcal{V}$ defined in Eq.~(\ref{eq:V_definition})
is a conserved quantity. Specifically, it is
\begin{equation}
\mathcal{V}=32B\vec{v}\cdot\vec{J},
\end{equation}
where $\vec{J}=(J_{x},J_{y},J_{z})^{\mathrm{T}}$ and $\vec{v}=(1,0,2B)^{\mathrm{T}}$.
Using this method, we find a non-trivial conserved quantity in this
two-spin system. Moreover, based on the property of quantum conserved
quantity, all the operators, for which the corresponding vectors share
the same or opposite directions with $\vec{v}$, are conserved quantities.
$\vec{v}$ is a vector in the $x-z$ plane. When $B$ is zero, $\vec{v}$
is along the $x$ axis. With the increase of $B$, $\vec{v}$ rotates
around the $y$ axis from $x$ axis to $z$ axis. For a very
large $B$, $\vec{v}$ is almost along the $z$ axis.

%----------------------Figure 1-------------------
\begin{figure}
\includegraphics[width=8cm]{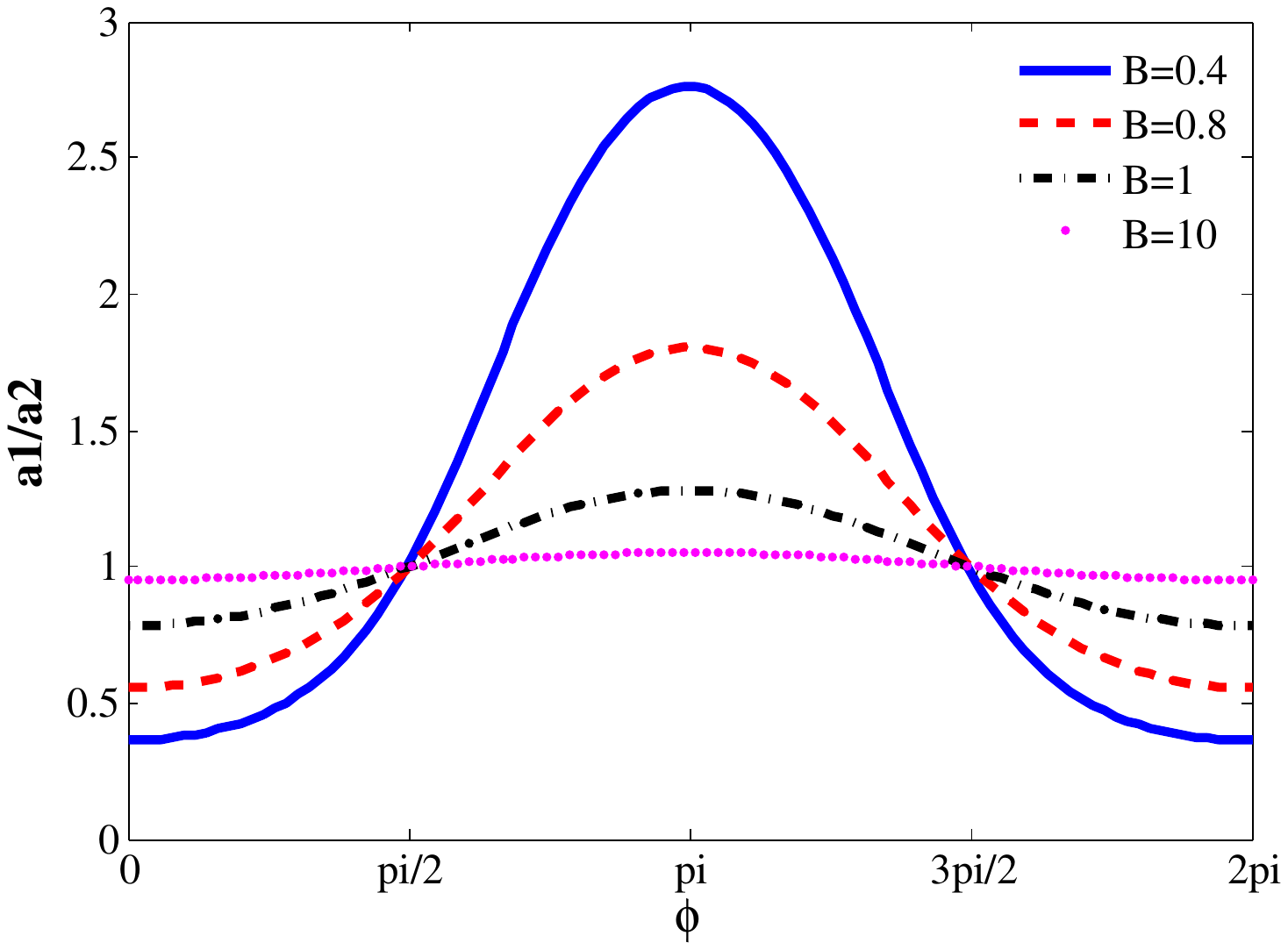}
\protect\caption{\label{fig:theta_a1a2}(Color online) Optimal points to access the
maximum quantum Fisher information for ferromagnetic two-spin system
at the long-time limit. The solid blue, dashed red, dash-dot black
and dotted pink lines represent the optimal points for $B=0.4,$ $0.8$,
$1.0$ and $10$, respectively.}
\end{figure}
%--------------------------------------------------

Compared with the expressions of $\Omega$ and $\vec{v}$, one can
see that $\Omega=2v$ with $v=|\vec{v}|$. the function $f$ can then
be rewritten into $f=\left[2vt-\sin\left(2vt\right)\right]/8v^{3}.$
The characteristic operator $\mathcal{H}_{B}$ can be calculated
via Eq.~(\ref{eq:H_theta}). Its explicit expression is
\begin{equation}
\mathcal{H}_{B}=4\vec{x}\cdot\vec{J},
\end{equation}
where the vector $\vec{x}$ reads $\vec{x}=(8Bf,-2\partial_{t}f,t-4f)^{\mathrm{T}}.$
With above $\mathcal{H}_{B}$, the QFI for purely initial states in
this case can be written as
\begin{equation}
F_{B}=\frac{16}{3}\left[|\vec{x}|^{2}\langle|\vec{J}|^{2}\rangle
-3\left(\vec{x}\cdot\langle\vec{J}\rangle\right)^{2}\right],\label{eq:F_1}
\end{equation}
where $|\vec{J}|^{2}=3(1+\sigma_{1}^{z}\sigma_{2}^{z})/8$. During
the calculation of $F_{B}$, the relation $\{J_{i},J_{j}\}=2\delta_{ij}J_{i}$
has been used. Above equation of QFI implies that the its maximum
is attained when $\langle|\vec{J}|^{2}\rangle$ is maximum and $\langle\vec{J}\rangle$
is vertical to $\vec{x}$. In a 4-dimensional Hilbert space, the maximum
value of $\langle\sigma_{1}^{z}\sigma_{2}^{z}\rangle$ is 1, which
indicates the formula of the maximum QFI must be
\begin{equation}
F_{B,\mathrm{max}}=4|\vec{x}|^{2},
\end{equation}
and the optimal initial state is required to be in the form
\begin{equation}
|\psi_{\mathrm{opt}}\rangle=a_{1}|00\rangle+a_{2}e^{i\phi}|11\rangle,
\label{eq:opt_1}
\end{equation}
where $a_{1,2}$ is a real number. To make $\vec{x}\cdot\langle\vec{J}\rangle=0$,
the amplitudes need to satisfy the equation
\begin{equation}
a_{1}a_{2}\left(x_{x}\cos\phi+x_{y}\sin\phi\right)+\frac{1}{2}
x_{z}\left(a_{1}^{2}-a_{2}^{2}\right)=0.
\end{equation}
Here $x_{i}$ $(i=x,y,z)$ is a element of $\vec{x}$. Since $x_{z}=t-4f$
is not always zero for $t>0$, then $a_{1}a_{2}=0$ cannot be a solution
of this equation, which means above equation can be further simplified
into
\begin{equation}
x_{x}\cos\phi+x_{y}\sin\phi+\frac{1}{2}x_{z}\left(\frac{a_{1}}
{a_{2}}-\frac{a_{2}}{a_{1}}\right)=0.\label{eq:optimal}
\end{equation}
All states satisfying this equation are available to access $F_{\theta,\mathrm{max}}$.
At the long-time limit, $f\simeq t/4v^{2}$ and $\partial_{t}f\ll f$,
then Eq.~(\ref{eq:optimal}) reduces to
\begin{equation}
4B\left[\cos\phi+B\left(\frac{a_{1}}{a_{2}}-\frac{a_{2}}{a_{1}}\right)\right]=0.
\label{eq:opt_longtime}
\end{equation}

Figure~\ref{fig:theta_a1a2} gives the optimal points $(\theta,a_{1}/a_{2})$
to access the maximum QFI for different $B$ at the long-time limit.
The solid blue, dashed red, dash-dot black and dotted pink lines in
this figure represent the optimal points for $B=0.4,$ $0.8$, $1.0$
and $10$, respectively. From this figure, it can be found that with
the increase of $B$, the curve becomes more flat. This behavior indicates
that for a strong external field, the maximum QFI is insensitive to
the relative phase $\phi$. This is actually due to the fact that
when $B$ is very large in above equation, $\cos\phi$ can be neglected
and the equation reduces to
\begin{equation}
\frac{a_{1}}{a_{2}}-\frac{a_{2}}{a_{1}}=0.
\end{equation}
One solution of this equation is $a_{1}=a_{2}$. Thus, for a strong
external field, the maximum QFI can be saturated at $a_{1}=a_{2}$
with any phase $\phi$.

\begin{figure}
\includegraphics[width=7cm]{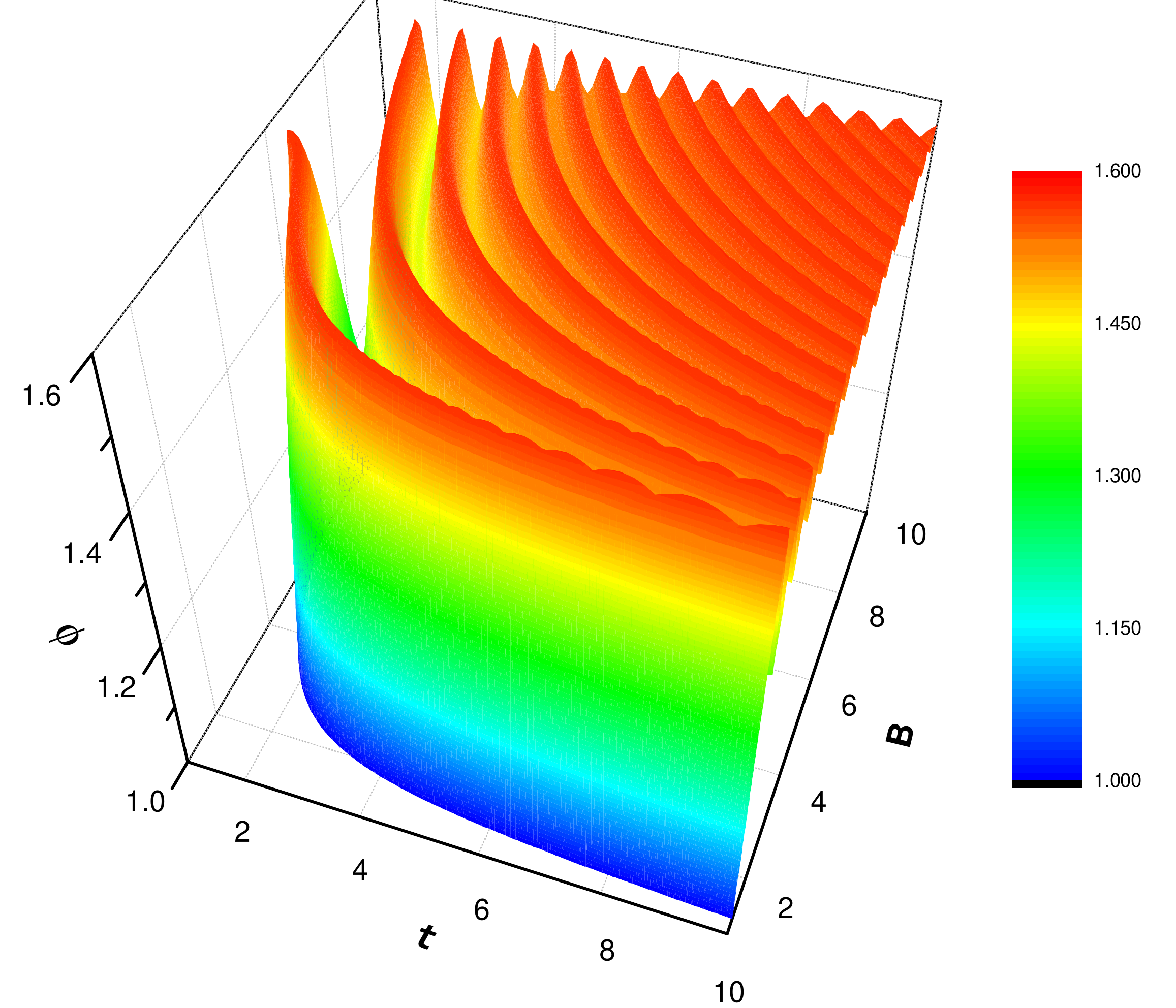}
\protect\caption{\label{fig:rela_phase}(Color online)
The optimal relative phase $\phi_{\mathrm{opt}}$
as a function of $B$ and $t$ for the optimal state $(|00\rangle+e^{i\phi}|11\rangle)/\sqrt{2}$. }
\end{figure}

A widely studied special form is that $a_{1}=a_{2}=1/\sqrt{2}$, i.e.,
$|\psi_{\mathrm{opt}}\rangle=(|00\rangle+e^{i\phi}|11\rangle)/\sqrt{2}$,
then the optimal relative phase $\phi_{\mathrm{opt}}$ reads
\begin{equation}
\phi_{\mathrm{opt}}=\arctan\left(\frac{4Bf}{\partial_{t}f}\right).
\end{equation}
At the long-time limit, this optimal phase reduces to a constant number
$\pi/2$, independent of the external field. This fact means at the long-time
limit, the state $(|00\rangle+e^{i\phi}|11\rangle)/\sqrt{2}$ is always
an optimal state for any strength of external field. Figure~\ref{fig:rela_phase}
shows the general variation of optimal relative phase as a function
of $B$ and $t$. For a weak external field, $\phi_{\mathrm{opt}}$
is growing rapidly with the increase of $B$ and $t$. This is because
when the external field is very weak, $4Bf/\partial_{t}f\simeq4Bt/3$,
then $\phi_{\mathrm{opt}}\simeq\arctan(4Bt/3)\simeq4Bt/3.$ Thus,
the optimal phase $\phi_{\mathrm{opt}}$ grows almost linearly with
$B$ and $t$ in this regime. With the continue increase of $B$ and
$t$, $\phi_{\mathrm{opt}}$ shows a oscillating behavior for intermediate
strength of external field. The oscillation amplitude of $\phi_{\mathrm{opt}}$
trends to shrink with the passage of time as $\phi_{\mathrm{opt}}=\pi/2$
when $t$ is infinite large.

%---------------Figure 2--------------------

%--------------------------------------------

\emph{Thermal state.}-Here we consider the thermal state of this ferromagnetic
two-spin system, of which the density matrix reads $\rho_{1}=\exp(-\beta H_{1})/Z_{1}$,
with $Z_{1}=\mathrm{Tr}\left[\exp(-\beta H_{1})\right]$. It is already
known that $\mathcal{V}$ is a conserved quantity, then based on Eq.~(\ref{eq:L_thermal}),
the SLD operator for $B$ of this thermal state can be written as
\begin{equation}
L_{B}=\beta r\mathcal{V}+\frac{2}{v}\tanh\left(2\beta v\right)J_{z}
\end{equation}
where $r$ is defined in Eq.~(\ref{eq:R}) and in this case, it has
the form
\begin{equation}
r=\frac{1}{4v^{2}}\left[1-\frac{1}{2\beta v}\tanh\left(2\beta v\right)\right].
\end{equation}

Based on the expressions of $L_{B}$, the QFI can be calculated as
\begin{equation}
F_{\mathrm{T}}=2\beta^{2}\left(16v^{2}r^{2}-8r+1\right)
\left(1+\langle\sigma_{1}^{z}\sigma_{2}^{z}\rangle\right).
\end{equation}
$F_{T}$ here is determined by the correlation
function $\langle\sigma_{1}^{z}\sigma_{2}^{z}\rangle$, which can
be analytically solved as
\begin{equation}
\langle\sigma_{1}^{z}\sigma_{2}^{z}\rangle=-1
+\frac{2\cosh\left(v\beta\right)}{\cosh\left(v\beta\right)+\cosh\beta}.
\end{equation}
Thus, $F_{\mathrm{T}}$ can be finally expressed in the form
\begin{equation}
F_{\mathrm{T}}=\left(16v^{2}r^{2}-8r+1\right)\frac{4\beta^{2}
\cosh\left(v\beta\right)}{\cosh\left(v\beta\right)+\cosh\beta}.
\label{eq:F_T}
\end{equation}

%--------------------Figure 3-----------------------
\begin{figure}
\includegraphics[width=7cm]{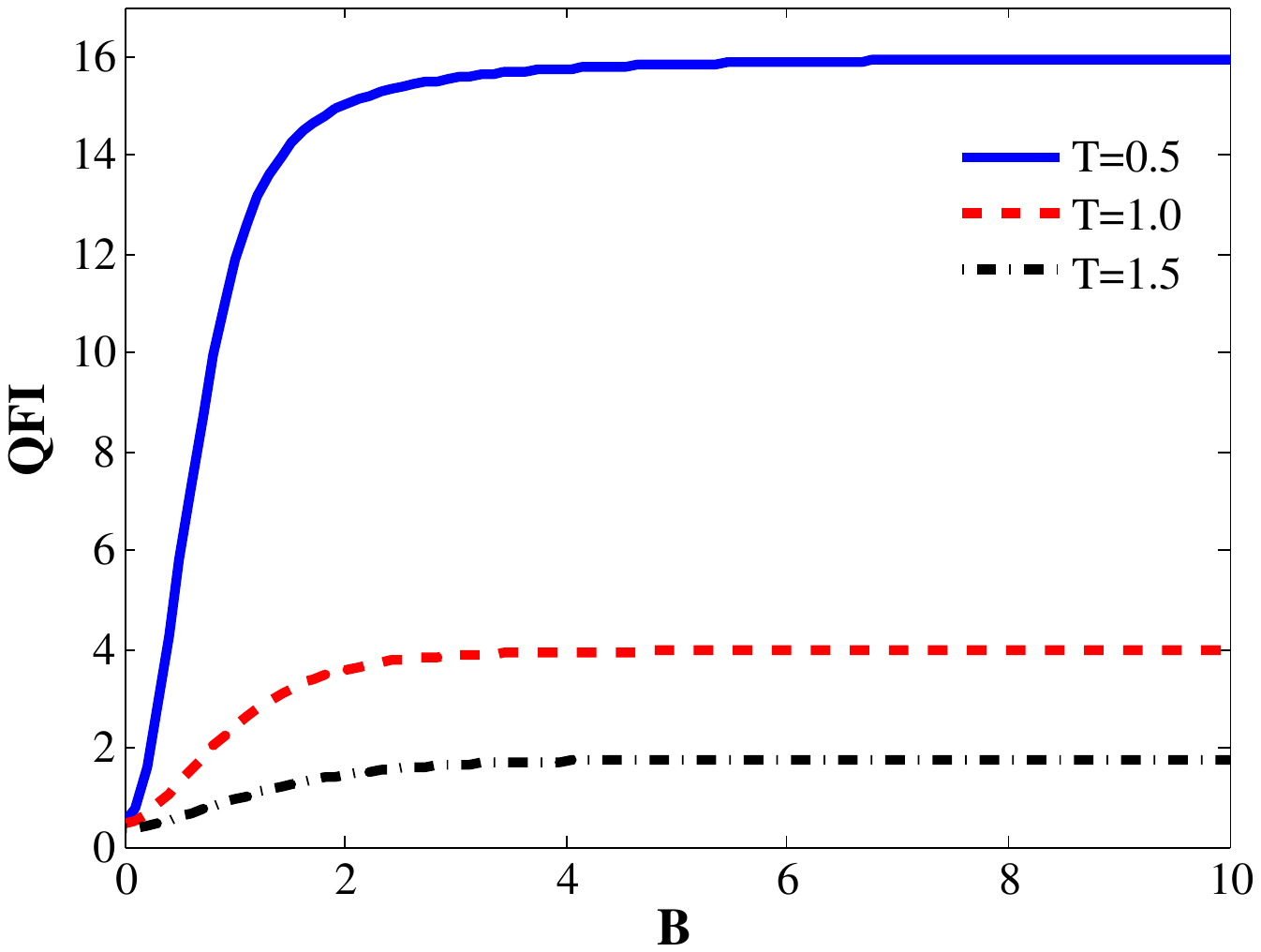}
\protect\caption{\label{fig:thermal}(Color online) Quantum Fisher
information as a function of~$B$~for thermal state of the
ferromagnetic two-spin system.
The temperatures are set as $T=0.5$, $1.0$ and $1.5$ for the solid
blue, dashed red and dash-dot black lines. }
\end{figure}
%---------------------------------------------------

Figure~\ref{fig:thermal} shows the variation of $F_{\mathrm{T}}$
as a function of $B$. The temperatures are set as $T=0.5$, $1.0$
and $1.5$ for the solid blue, dashed red and dash-dot black lines.
In this figure, for the weak external field, the QFI grows greatly
with the increase of $B$. However, for the strong external field,
this growth is not significant. With respective to the temperature,
the decrease of $T$ shows a positive effect on the QFI and is useful
for the precision measure in this system.

For a very low temperature and nonzero external field, $\cosh\beta/\cosh(v\beta)\simeq e^{\beta(1-v)}\simeq0$,
then the QFI in Eq.~(\ref{eq:F_T}) reduces to the form
\begin{equation}
F_{\mathrm{T}}\simeq\frac{4}{T^{2}}\left(1-\frac{1}{v^{2}}\right).
\end{equation}
This equations shows that in the low-temperature regime, the QFI is
inversely proportional to the square of $T$. Thus, the decrease of
$T$ will dramatically improve the value of $F_{\mathrm{T}}$. Meanwhile,
the increase of $B$ will also enhance the value of QFI. The maximum
value of $F_{\mathrm{T}}$ in this regime is $F_{\mathrm{T,max}}=4/T^{2}$.

\subsubsection*{Anisotropic two-spin system}

In the following we consider a more general case: the anisotropic
two-spin ferromagnetic XY model with an inhomogeneous external magnetic
field. The Hamiltonian of this system is
\begin{eqnarray}
H_{2} &=& -\frac{1+\gamma}{2}\sigma_{1}^{x}\sigma_{2}^{x}
-\frac{1-\gamma}{2}\sigma_{1}^{y}\sigma_{2}^{y}\nonumber \\
& & -B_{+}\left(\sigma_{1}^{z}+\sigma_{2}^{z}\right)
-B_{-}\left(\sigma_{1}^{z}-\sigma_{2}^{z}\right),\label{eq:H_2}
\end{eqnarray}
where $\gamma$ is the anisotropic parameter. This Hamiltonian can
reduces to the Hamiltonian $H_{1}$ in Eq.~(\ref{eq:H_1}) with $\gamma=1$
and $B_{-}=0$. Before the main calculation, we introduce a new group
of operators
\begin{equation}
\begin{cases}
S_{x}=\!\! & \frac{1}{4}\left(\sigma_{1}^{x}\sigma_{2}^{y}
-\sigma_{1}^{y}\sigma_{2}^{x}\right),\\
S_{y}=\!\! & \frac{1}{4}\left(\sigma_{1}^{x}\sigma_{2}^{x}
+\sigma_{1}^{y}\sigma_{2}^{y}\right),\\
S_{z}=\!\! & \frac{1}{4}\left(\sigma_{1}^{z}-\sigma_{2}^{z}\right).
\end{cases}
\end{equation}
Similarly with $J_{x,y,z}$, $S_{x,y,z}$ also satisfies $\mathfrak{su}(2)$
commutation, i.e., $[S_{i},S_{j}]=i\epsilon_{ijk}S_{k}$. Meanwhile,
the anti-commutation relation is $\{S_{i},S_{j}\}=2\delta_{ij}S_{i}$.
A very interesting property between these two groups of operators
is that
\begin{equation}
J_{i}S_{j}=S_{j}J_{i}=0,\quad\forall i,j=x,y,z.\label{eq:temp1}
\end{equation}
$[J_{i},S_{j}]=0$ for any $i$ and $j$ is a natural result of this
property.

Utilizing these two set of operators, Hamiltonian~(\ref{eq:H_2})
can be written as the sum of two parts, i.e.,
\begin{equation}
H_{2}=2(H_{+}+H_{-}),
\end{equation}
where the sub-Hamiltonians $H_{+}$ is only related to $J_{x,y,z}$
and $H_{-}$ is only related to $S_{x,y,z}$. Their specific formulas are
\begin{eqnarray}
H_{+} & = & -\gamma J_{x}-2B_{+}J_{z},\\
H_{-} & = & -S_{y}-2B_{-}S_{z}.
\end{eqnarray}
Since $J_{i}$ and $S_{j}$ are commutative for any $i$ and $j$,
$H_{+}$ and $H_{-}$ are also commutative. Here we take both $B_{+}$
and $B_{-}$ as the parameters under estimation. Through some algebra,
we find that if one choose
\begin{equation}
\Omega_{+}=2\sqrt{\gamma^{2}+4B_{+}^{2}},\quad\Omega_{-}=2\sqrt{1+4B_{-}^{2}},
\end{equation}
the corresponding operator $\mathcal{V}_{+}$ and $\mathcal{V}_{-}$
defined in Eq.~(\ref{eq:V_definition}) are conserved quantities
and have the form
\begin{equation}
\mathcal{V}_{+}=32B_{+}\vec{v}_{+}\cdot\vec{J},\quad\mathcal{V}_{-}
=32B_{-}\vec{v}_{-}\cdot\vec{S},
\end{equation}
where $\vec{J}=(J_{x},J_{y},J_{z})^{\mathrm{T}}$, $\vec{S}=(S_{x},S_{y},S_{z})^{\mathrm{T}}$
and the vector $\vec{v}_{+}$ and $\vec{v}_{-}$ read
\begin{equation}
\vec{v}_{+}=\left(\gamma,0,2B_{+}\right)^{\mathrm{T}},\quad\vec{v}_{-}
=\left(0,1,2B_{-}\right)^{\mathrm{T}}.
\end{equation}
Then there is $\Omega_{\pm}=2v_{\pm}$ with $v_{\pm}=|\vec{v}_{\pm}|$.

\begin{figure}
\includegraphics[width=7cm]{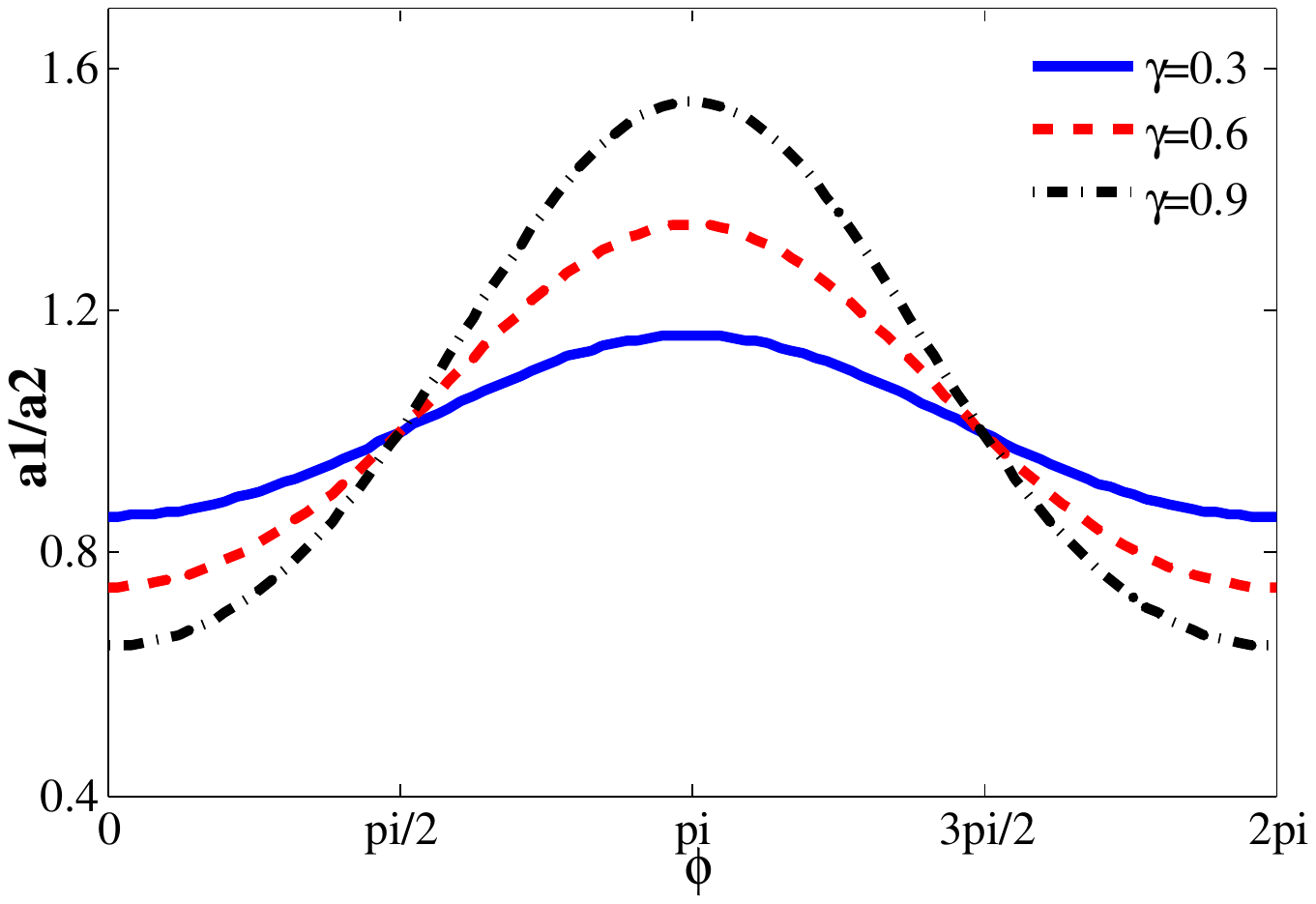}
\protect\caption{\label{fig:anisotropic}(Color online) Optimal
points to access the maximum quantum Fisher information for
anisotropic two-spin system at the long-time limit. The solid blue,
dashed red, and dash-dot black lines represent the optimal points
for $\gamma=0.3$, $0.6$ and $0.9$. }
\end{figure}

Utilizing these conserved quantities, the characteristic operators
$\mathcal{H}_{+}$ and $\mathcal{H}_{-}$ can be expressed by
\begin{equation}
\mathcal{H}_{+}=4\vec{x}_{+}\cdot\vec{J},\quad\mathcal{H}_{-}
=4\vec{x}_{-}\cdot\vec{S}.\label{eq:H_pm}
\end{equation}
The vector $\vec{x}_{\pm}$ in above equation reads
\begin{eqnarray}
\vec{x}_{+} & = & \left(8\gamma B_{+}f_{+},-2\gamma\partial_{t}f_{+},t-4\gamma^{2}f_{+}\right)^{\mathrm{T}},\\
\vec{x}_{-} & = & \left(2\partial_{t}f_{-},8B_{-}f_{-},t-4f_{-}\right)^{\mathrm{T}}.
\end{eqnarray}
Here $f_{\pm}=\left[2v_{\pm}t-\sin\left(2v_{\pm}t\right)\right]/8\vec{v}_{\pm}^{3}$.
For purely initial states, the QFI is the variance of $\mathcal{H}$
on the initial state. For parameter $B_{+}$, the QFI $F_{+}$ shares
the same form with that in Eq.~(\ref{eq:F_1}), i.e.,
\begin{equation}
F_{+}=\frac{16}{3}\left[|\vec{x}_{+}|^{2}\langle|\vec{J}|^{2}\rangle
-3\left(\vec{x}_{+}\cdot\langle\vec{J}\rangle\right)^{2}\right].
\end{equation}
For parameter $B_{-}$, the QFI $F_{-}$ has the similar form
\begin{equation}
F_{-}=\frac{16}{3}\left[|\vec{x}_{-}|^{2}\langle|\vec{S}|^{2}\rangle
-3\left(\vec{x}_{-}\cdot\langle\vec{S}\rangle\right)^{2}\right],
\end{equation}
where $|\vec{S}|^{2}=3(1-\sigma_{1}^{z}\sigma_{2}^{z})/8$. The maximum
value of $F_{+}$ is $4|\vec{x}_{+}|^{2}$ and the corresponding optimal
initial state has the same form as that in Eq.~(\ref{eq:opt_1}).
For the long-time limit, the equation for $B_{+}$ that the optimal
points satisfy is
\begin{equation}
4\gamma B_{+}\cos\phi+\left(v_{+}^{2}-\gamma^{2}\right)\left(\frac{a_{1}}{a_{2}}
-\frac{a_{2}}{a_{1}}\right)=0.
\end{equation}
Figure~\ref{fig:anisotropic} gives the optimal points to access
the maximum QFI for the long-time limit. The solid blue, dashed red,
and dash-dot black lines represent the optimal points for $\gamma=0.3$,
$0.6$ and $0.9$. It is found in this figure that with the increase
of $\gamma$, the curve of the optimal points gets more sharp, indicating
that the maximum QFI is more sensitive to the relative phase for a
large $\gamma$.

%----------------------Figure 4-------------------
%-------------------------------------------------

Similarly, the maximum value of $F_{-}$ is $4|\vec{x}_{-}|^{2}$
with the optimal initial state
\begin{equation}
|\Phi_{\mathrm{opt}}\rangle=b_{1}|01\rangle+b_{2}e^{i\varphi}|10\rangle,
\end{equation}
where $b_{1}$ and $b_{2}$ are real numbers. The equation for $b_{1}$,
$b_{2}$ to satisfy to access the maximum QFI is
\begin{equation}
2\partial_{t}f_{-}\sin\varphi-8B_{-}f_{-}\cos\varphi=\frac{1}{2}\!\left(t-4f_{-}\right)\!\!\left(\!\frac{b_{1}}{b_{2}}-\frac{b_{2}}{b_{1}}\!\right)\!.
\end{equation}
For the long-time limit, this equation reduces to the same form with
Eq.~(\ref{eq:opt_longtime}). Moreover, taking $b_{1}=b_{2}=1/\sqrt{2}$,
the optimal relative phase can be written as
\begin{equation}
\varphi_{\mathrm{opt}}=\arctan\left(\frac{4B_{-}f_{-}}{\partial_{t}f_{-}}\right).
\end{equation}
At the long-time limit, $\varphi_{\mathrm{opt}}$ also equals to $\pi/2$.

Now we consider the situation that both $B_{+}$ and $B_{-}$ are
unknown parameters simultaneously. It is known the condition for the
saturation of the multiparameter Cram\'{e}r-Rao bound in unitary
parametrization process is~\cite{JLiu_H}
\begin{equation}
\langle\psi_{\mathrm{in}}|\left[\mathcal{H}_{+},\mathcal{H}_{-}\right]
|\psi_{\mathrm{in}}\rangle=0.
\end{equation}
From the expressions of $\mathcal{H}_{\pm}$ in Eq.~(\ref{eq:H_pm})
and the property that $J_{i}$ and $S_{j}$ are commutative for all
$i$ and $j$, it is easy to see that $[\mathcal{H}_{+},\mathcal{H}_{-}]=0$.
Thus, $B_{+}$ and $B_{-}$ can be jointly measured for any purely
initial state. Since $\mathcal{H}_{+}$ and $\mathcal{H}_{-}$ are
commutative and based on Eq.~(\ref{eq:temp1}), the off-diagonal
element of QFIM is
\begin{equation}
\mathcal{F}_{+-}=\mathcal{F}_{-+}=-\langle\mathcal{H}_{+}\rangle
\langle\mathcal{H}_{-}\rangle.
\end{equation}
The expected value above is taken on the initial state. According
to the Cram\'{e}r-Rao theory, we have
\begin{equation}
\delta^{2}B_{\pm}\geq\frac{\mathcal{F}_{\mp}\mathcal{F}_{\mp}}
{\mathcal{F}_{++}\mathcal{F}_{--}-\mathcal{F}_{+-}^{2}}.
\end{equation}

When the initial state is chosen as $|\psi_{\mathrm{opt}}\rangle$
or $|\Phi_{\mathrm{opt}}\rangle$, $\mathcal{F}_{+-}$ vanishes and
the inequality above reduces to the form of the single-parameter cases,
which indicates that the joint measurement of $B_{+}$ and $B_{-}$
can be performed by $|\psi_{\mathrm{opt}}\rangle$ or $|\Phi_{\mathrm{opt}}\rangle$.
However, it should be noticed that $|\psi_{\mathrm{opt}}\rangle$
and $|\Phi_{\mathrm{opt}}\rangle$ are orthogonal, which means even
$B_{+}$ and $B_{-}$ can be jointly measured, there does not exist
an optimal state to access the maximum QFI for both $B_{+}$ and $B_{-}$ 
simultaneously. This fact implies that the joint measurement here
is not as good as the single measurement.

\emph{Thermal state}.-Next we consider the thermal state of $H_{2}$,
which is $\rho_{2}=\exp(-\beta H_{2})/Z_{2}$, with $Z_{2}$ the partition
function. Based on Eq.~(\ref{eq:L_thermal}), the SLD operators for
$B_{+}$ and $B_{-}$ can be expressed by
\begin{equation}
L_{\pm}=\beta r_{\pm}\mathcal{V}_{\pm}-\frac{1}{v_{\pm}}\tanh\left(2\beta v_{\pm}\right)\partial_{\pm}H_{\pm}.
\end{equation}
where $\partial_{\pm}$ is short for $\partial_{B_{\pm}}$ and the
coefficient
\begin{equation}
r_{\pm}=\frac{1}{4v_{\pm}^{2}}\left[1-\frac{1}{2\beta v_{\pm}}\tanh\left(2\beta v_{\pm}\right)\right].
\end{equation}
Since $\{S_{i},S_{j}\}=\{J_{i},J_{j}\}=0$ for $i\neq j$, the thermal
QFI for $B_{+}$ and $B_{-}$ can then be written as
\begin{eqnarray}
F_{\mathrm{T+}} & \!\!=\!\! & 2\beta^{2}\!\left(16\gamma^{2}v_{+}^{2}r_{+}^{2}\!-\!8\gamma^{2}
r_{+}\!+\!1\right)\!\left(1\!+\!\langle\sigma_{1}^{z}\sigma_{2}^{z}
\rangle\right)\!,\\
F_{\mathrm{T-}} & \!\!=\!\! & 2\beta^{2}\left(16v_{-}^{2}r_{-}^{2}-8r_{-}+1\right)
\left(1-\langle\sigma_{1}^{z}\sigma_{2}^{z}\rangle\right).
\end{eqnarray}
For the thermal state, the correlation function $\langle\sigma_{1}^{z}\sigma_{2}^{z}\rangle$ is
\begin{equation}
\langle\sigma_{1}^{z}\sigma_{2}^{z}\rangle=-1+\frac{2\cosh\left(\beta v_{+}\right)}{\cosh\left(\beta v_{+}\right)+\cosh\left(\beta v_{-}\right)}.
\end{equation}

In the low-temperature regime, there is
\begin{equation}
\frac{\cosh(\beta v_{-})}{\cosh(\beta v_{+})}\simeq e^{\beta(v_{-}-v_{+})}.
\end{equation}
Therefore, when $v_{+}$ equals to $v_{-}$, $e^{\beta(v_{-}-v_{+})}$
equals to 1, and $\langle\sigma_{1}^{z}\sigma_{2}^{z}\rangle=0$.
$F_{\mathrm{T},+}$ and $F_{\mathrm{T},-}$ reduce to
\begin{equation}
F_{\mathrm{T},+}\simeq\frac{2}{T^{2}}
\left(1-\frac{\gamma^{2}}{v_{+}^{2}}\right)\!,\quad F_{\mathrm{T},-}\simeq\frac{2}{T^{2}}
\left(1-\frac{1}{v_{-}^{2}}\right)\!.
\end{equation}
When $v_{+}$ is smaller than $v_{-}$, $e^{\beta(v_{-}-v_{+})}$
trends to infinity in the low-temperature regime, then $\langle\sigma_{1}^{z}\sigma_{2}^{z}\rangle\simeq-1$
and $F_{\mathrm{T},+}$ and $F_{\mathrm{T},-}$ is in the form
\begin{equation}
F_{\mathrm{T,+}}\simeq0,\quad F_{\mathrm{T,-}}\simeq\frac{4}{T^{2}}\left(1-\frac{1}{v_{-}^{2}}\right).
\end{equation}
When $v_{+}$ is larger than $v_{-}$, $e^{\beta(v_{-}-v_{+})}\simeq0$,
then $\langle\sigma_{1}^{z}\sigma_{2}^{z}\rangle\simeq1$, and the
QFI $F_{\mathrm{T},+}$ and $F_{\mathrm{T},-}$ can be written as
\begin{equation}
F_{\mathrm{T},+}\simeq\frac{4}{T^{2}}
\left(1-\frac{\gamma^{2}}{v_{+}^{2}}\right)\!,\quad F_{\mathrm{T},-}\simeq0.
\end{equation}
Above analysis shows that when $v_{+}$ is smaller than $v_{-}$,
the parameter $B_{+}$ can be barely estimated via the Cram\'{e}r-Rao
theory, so as $B_{-}$ when $v_{+}$ is larger than $v_{-}$. Thus,
if either of $B_{+}$ and $B_{-}$ is the parameter under estimation,
we have to tune down the value of the other one to make sure that 
nonzero QFI exists.

\subsubsection*{Spin-one model}

Not only the spin-half systems, but also some spin-one systems, can
fit in the $\mathfrak{su}(2)$ category. Here we show such a spin-one
system in the one-axis twisting model. The Hamiltonian of a one-axis
twisting model with a transverse field can be written in the form~\cite{Sanders89,Kitagawa93,Law2001,Rojo2003}
\begin{equation}
H_{3}=\chi J_{x}^{2}+BJ_{z},
\end{equation}
where $J_{x}=(a^{\dagger}b+ab^{\dagger})/2$ and $J_{z}=(a^{\dagger}a-b^{\dagger}b)/2$
are the Schwinger operators. Another one is $J_{y}=(a^{\dagger}b-ab^{\dagger})/2i$.
$\chi$ is the coupling constant and $B$ is the strength of the transverse
field. This Hamiltonian can be realized in many physical systems including
two-component Bose-Einstein condensates~\cite{sorensen2001,sorensen2002}.
Now we consider one realization that a two-boson system in a double
well. Since the particle number is a conserved quantity, this system
can be expanded in the basis $\{|02\rangle,|11\rangle,|20\rangle\}$
in Fock space, which can be mapped as a spin-one system. In this basis,
the Schwinger operators has the form
\begin{equation}
J_{x}=\frac{1}{\sqrt{2}}\left(\begin{array}{ccc}
0 & 1 & 0\\
1 & 0 & 1\\
0 & 1 & 0
\end{array}\right)\!\!,\ J_{y}=\frac{i}{\sqrt{2}}\left(\begin{array}{ccc}
0 & 1 & 0\\
-1 & 0 & 1\\
0 & -1 & 0
\end{array}\right)\!\!,
\end{equation}
and
\begin{equation}
J_{z}=\left(\begin{array}{ccc}
-1 & 0 & 0\\
0 & 0 & 0\\
0 & 0 & 1
\end{array}\right).
\end{equation}
Based on these matrices and taking the parameter $B$ as the one under
estimation, one can easily check that the Hamiltonian $H_{3}$ satisfies
the following equation
\begin{equation}
[(H_{3}^{\times})^{2}-(\chi^{2}+4B^{2})]H_{3}^{\times}\partial_{\theta}H_{3}=0,
\end{equation}
which implies that if we choose $\Omega=\sqrt{\chi^{2}+4B^{2}},$
the operator $\mathcal{V}$ defined in Eq.~(\ref{eq:V_definition})
will be a conserved quantity. Specifically, $\mathcal{V}$ can be
written in the form
\begin{equation}
\mathcal{V}=\left(\chi^{2}-\Omega^{2}\right)J_{z}-2B\chi\mathcal{I},
\end{equation}
where $\mathcal{I}=|02\rangle\langle20|+|20\rangle\langle02|.$ Furthermore,
the characteristic function $\mathcal{H}_{B}$ can be written as
\begin{equation}
\mathcal{H}_{B}=\left(\chi^{2}f-t\right)J_{z}-2B\chi\mathcal{I}+\frac{i2}{\Omega^{2}}\sin^{2}\left(\frac{\Omega}{2}t\right)J_{z}\mathcal{I}.
\end{equation}
For the long-time limit, it reduces to
\begin{equation}
\mathcal{H}_{B}=\frac{-4B^{2}t}{\chi^{2}+4B^{2}}J_{z}-2B\chi\mathcal{I}.
\end{equation}
For the unitary parametrization processes with a purely initial state,
the QFI for above $\mathcal{H}_{B}$ reads
\begin{equation}
F_{B}=\left(\frac{4B^{2}t}{\chi^{2}+4B^{2}}\right)^{\!2}
\!\langle\Delta^{2}J_{z}\rangle+4B^{2}\chi^{2}\langle\Delta^{2}
\mathcal{I}\rangle,
\end{equation}
where we have used the equality $\{J_{z},\mathcal{I}\}=0.$ Furthermore,
the QFI can be simplified into
\begin{eqnarray}
F_{B} & = & \left[\left(\frac{4B^{2}t}{\chi^{2}+4B^{2}}\right)^{\!2}\!\!
+4B^{2}\chi^{2}\right]\langle J_{z}^{2}\rangle\nonumber \\
&  & -\left(\frac{4B^{2}t}{\chi^{2}+4B^{2}}\right)^{\!2}\!\langle J_{z}\rangle^{2}-4B^{2}\chi^{2}\langle\mathcal{I}\rangle^{2}.
\end{eqnarray}
During the calculation we used the equality $\langle J_{z}^{2}\rangle=\langle\mathcal{I}^{2}\rangle$.
The maximum value of QFI above is attained when $\langle J_{z}^{2}\rangle$
reaches its maximum value and $\langle J_{z}\rangle$, $\langle\mathcal{I}\rangle$
vanish simultaneously. Denoting the initial state as $|\psi\rangle=c_{1}|02\rangle+c_{2}e^{i\phi_{2}}|11\rangle+c_{3}
e^{i\phi_{3}}|20\rangle$~with $c_{1,2,3}$ real numbers, there are $\langle J_{z}^{2}\rangle=\langle\mathcal{I}^{2}\rangle=c_{1}^{2}+c_{3}^{2}$,
$\langle J_{z}\rangle=c_{1}^{2}-c_{3}^{2}$ and $\langle\mathcal{I}\rangle=2c_{1}c_{3}\cos\phi_{3}$.
Utilizing these expressions, it can be checked that when $c_{1}=c_{3}=1/\sqrt{2}$
and $\phi_{3}=\pi/2$, all the conditions can be satisfied simultaneously,
which implies that one optimal initial state here is a NOON-type state,
i.e.,
\begin{equation}
|\psi_{\mathrm{opt}}\rangle=\frac{1}{\sqrt{2}}\left(|02\rangle+i|20\rangle\right).
\end{equation}
The corresponding maximum QFI is
\begin{equation}
F_{B,\mathrm{max}}=4B^{2}\left[\frac{4B^{2}t^{2}}
{\left(\chi^{2}+4B^{2}\right)^{2}}+\chi^{2}\right].
\end{equation}
This expression shows that the maximum QFI will be square-enhanced
with the passage of time. For the coupling constant $\chi$, $F_{B,\mathrm{max}}$
does not change monotonously. For a small $\chi$, $F_{B,\mathrm{max}}$
increases sharply with the decrease of $\chi$ and the trend is totally
reverse for a large $\chi$. The minimum value of $F_{B,\mathrm{max}}$
is attained around $\chi^{2}=4(Bt)^{2/3}-4B^{2}$. Thus, the value
of $\chi$ should be tuned carefully to avoid this regime during the
measure of the transverse field.

\subsection{Canonical category}

%-----------------------Figure 5--------------------
\begin{figure}
\includegraphics[width=7cm]{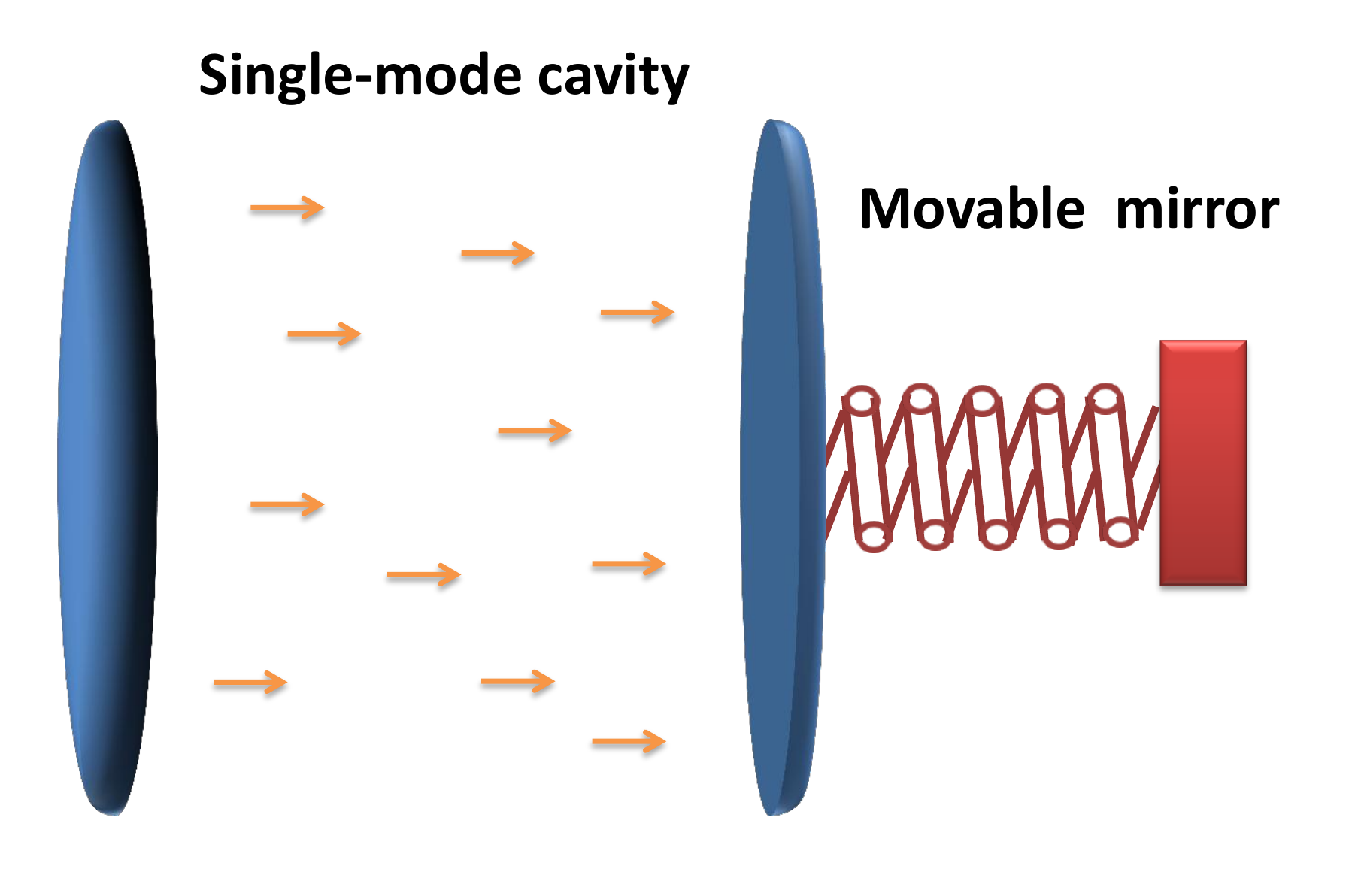}
\protect\caption{\label{fig:optomechanics} The schematic of a cavity optomechanics
system. The left device is a single-mode cavity and the right one
is a movable mirror.}
\end{figure}
%---------------------------------------------------

The most obvious property of Hamiltonians in canonical category is
that they can be partly or entirely rewritten via the canonical variables:
$x$, $p$ and the number operator. A typical case of canonical category
is the optomechanical systems, which has been widely discussed as
a novel artificial device~\cite{RMP_opto}. A simple model for optomechanical
systems is a single-mode cavity coupling with a movable mirror, of
which the schematic is shown in Fig.~\ref{fig:optomechanics}. The
total Hamiltonian can be written as~\cite{Bose,CKLaw}
\begin{equation}
H=\omega_{a}a^{\dagger}a+\omega_{b}b^{\dagger}b-\bar{g}a^{\dagger}a
\left(b+b^{\dagger}\right),\label{eq:H_opme}
\end{equation}
where $a$, $a^{\dagger}$, $b$, $b^{\dagger}$ are the annihilation
and creation operators for the cavity and mirror, respectively. $\omega_{a}$,
$\omega_{b}$ are the corresponding frequencies. The coupling strength
$\bar{g}$ has the form
\begin{equation}
\bar{g}=\frac{\omega_{a}}{l}\sqrt{\frac{1}{2m\omega_{b}}},
\end{equation}
with $l$, $m$ the length of the cavity and mass of the mirror. With
the introduction of the quadratic operators $x_{a(b)}$, $p_{a(b)}$
defined as
\begin{eqnarray}
x_{a} & = & \frac{1}{\sqrt{2}}\left(a+a^{\dagger}\right),\quad p_{a}=\frac{1}{\sqrt{2}i}\left(a-a^{\dagger}\right),\\
x_{b} & = & \frac{1}{\sqrt{2}}\left(b+b^{\dagger}\right),\quad p_{b}=\frac{1}{\sqrt{2}i}\left(b-b^{\dagger}\right),
\end{eqnarray}
and the number operators $N_{a}=a^{\dagger}a$, $N_{b}=b^{\dagger}b$,
Hamiltonian~(\ref{eq:H_opme}) can be rewritten into the form
\begin{equation}
H=\omega_{a}N_{a}+\omega_{b}N_{b}-gN_{a}x_{b},
\end{equation}
where $g=\sqrt{2}\bar{g}$. In this case we take $m$ or $l$ as the
parameter under estimation. Utilizing the commutation relations $[N,x]=-ip$,
$[N,p]=ix$ and $[x,p]=i$, one can see that
\begin{eqnarray}
H^{\times}\partial_{m(l)}H & = & i\omega_{b}g^{\prime}N_{a}p_{b},\\
\left(H^{\times}\right)^{2}\partial_{m(l)}H & = & -\omega_{b}^{2}g^{\prime}N_{a}x_{b}+\omega_{b}gg^{\prime}N_{a}^{2},
\end{eqnarray}
where we denote $g^{\prime}:=\partial_{m(l)}g$. From these equations,
one can check that if we choose $\Omega=\omega_{b}$, the operator
$\mathcal{V}$ defined in Eq.~(\ref{eq:V_definition}) is a conserved
quantity. Its specific expression is
\begin{equation}
\mathcal{V}=\omega_{b}gg^{\prime}N_{a}^{2}.
\end{equation}
As a matter of fact, the photon number in the cavity, i.e., $N_{a}$ 
is a conserve quantity, thus, it is natural
that any exponentiation of $N_{a}$ is also a conserved quantity.
Based on above information, the characteristic function reads
\begin{equation}
\mathcal{H}_{m(l)}=g^{\prime}N_{a}\left[\left(t-\omega_{b}^{2}f\right)
x_{b}+\omega_{b}(\partial_{t}f)p_{b}+g\omega_{b}fN_{a}\right].
\end{equation}
In the Fock space of the cavity, $N_{a}$ in above equation can be
replaced by the average photon number $n_{a}$, which is a constant, therefore,
the characteristic operator can be simplified into
\begin{equation}
\mathcal{H}_{m(l)}=\frac{n_{a}g^{\prime}}{\omega_{b}}\left\{ \sin\left(\omega_{b}t\right)x_{b}+\left[1-\cos\left(\omega_{b}t\right)
\right]p_{b}\right\} .
\end{equation}
For a purely initial state, the QFI is in the form
\begin{eqnarray}
F_{m(l)} & \!\!\!=\!\!\! & \left(\!\frac{n_{a}g^{\prime}}{\omega_{b}}\!\right)^{\!\!2}
\!\!\!\big\{\!\sin^{2}\!\left(\omega_{b}t\right)
\!\langle\Delta^{2}x_{b}\rangle\!
+\!\left[1-\cos\left(\omega_{b}t\right)\right]^{2}
\!\langle\Delta^{2}p_{b}\rangle\nonumber \\
&  & +2\sin\left(\omega_{b}t\right)\left[1-\cos\left(\omega_{b}t\right)\right]
\mathrm{cov}\left(x_{b},p_{b}\right)\!\big\}.
\end{eqnarray}
If the movable mirror is initially in the vacuum state, above expression
of QFI reduces to
\begin{equation}
F_{m(l)}=\left(\!\frac{n_{a}g^{\prime}}{\omega_{b}}\!\right)^{\!\!2}
\left[1-\cos\left(\omega_{b}t\right)\right].
\end{equation}
When $t=(2k+1)\pi/\omega_{b}$ with $k=0,1,2,...,$ the QFI above
reaches its maximum value with respect to time, which is $F_{m(l),\mathrm{max}}=(n_{a}g^{\prime}/\omega_{b})^{2}$.
Contrarily, when $t=2k\pi/\omega_{b}$, the QFI vanishes, the parameter
cannot be estimated via Cram\'{e}r-Rao inequality. This fact shows that the
measure should not be performed at these time points. Moreover, the
increase of photon number in the cavity can squarely benefit the estimation
of $l$ and $m$.

For the mass $m$ and the length $l$, the specific expressions of
maximum QFI are
\begin{equation}
F_{m,\mathrm{max}}=\frac{n_{a}^{2}\omega_{a}^{2}}{4m^{3}l^{2}\omega_{b}^{5}},\quad F_{l,\mathrm{max}}=\frac{n_{a}^{2}\omega_{a}^{2}}{ml^{4}\omega_{b}^{3}}.
\end{equation}
In both expressions above, tuning down the frequency $\omega_{b}$
will help to improve the precision of $l$ and $m$. Especially for
the estimation of mass $m$, the decrease of $\omega_{b}$ will show
a dramatic enhancement of the precision.

There are several other systems in this category, including a quantum
harmonic oscillator in a classical field. The corresponding Hamiltonian
is $H=\omega_{\mathrm{ho}}a^{\dagger}a+ga^{\dagger}+g^{*}a$. In this
case, the characteristic operator $\mathcal{H}$ is also the linear
combination of operators $x$ and $p$. The $\mathfrak{su}(2)$ and
canonical categories discussed above are representative. 
However, there are still systems out of  these two categories in which $\mathcal{V}$ is a conserved quantity. For instance, a two-level atom in a single-mode cavity with the Hamiltonian $H=\omega_{a}a^{\dagger}a+\frac{1}{2}\omega_{0}\sigma_{z}
-g\left(a+a^{\dagger}\right)\sigma_{z}$~can also fit in the scenario discussed in this paper.

\section{Alternative form of QFI\label{sec:Alternative}}

The classical Fisher information has more than one extensions in quantum
mechanics. Besides the traditional one discussed above, an alternative
definition of quantum Fisher information is~\cite{SLuo}
\begin{equation}
I_{\theta}=4\mathrm{Tr}\left(\partial_{\theta}\sqrt{\rho}\right)^{2}.
\end{equation}
For the unitary parametrization $\rho(\theta)=U(\theta)\rho_{0}U^{\dagger}(\theta)$,
this alternative form of QFI can be expressed by
\begin{equation}
I_{\theta}=8\mathrm{Tr}\left[\mathcal{H}^{2}\rho_{0}
-\left(\mathcal{H}\sqrt{\rho_{0}}\right)^{2}\right],\label{eq:I}
\end{equation}
where $\mathcal{H}$ is the corresponding characteristic operator.
Similarly with the traditional expression, above formula is also determined
by $\mathcal{H}$ and the initial state $\rho_{0}$. Recalling the
spectral decomposition of $\rho_{0}$ as $\rho_{0}=\sum_{i=1}^{M}p_{i}|\psi_{i}\rangle\langle\psi_{i}|,$
with $M$ the dimension of the support of $\rho_{0}$, Eq.~(\ref{eq:I})
can be written into
\begin{equation}
I_{\theta}=8\sum_{i=1}^{M}\!\left(\! p_{i}\langle\Delta^{2}\mathcal{H}\rangle_{i}-2\sum_{j>i}^{M}
\sqrt{p_{i}p_{j}}|\langle\psi_{i}|\mathcal{H}|\psi_{j}\rangle|^{2}\!\right)\!,
\end{equation}
with $\langle\Delta^{2}\mathcal{H}\rangle_{i}$ the variance of $\mathcal{H}$
on the $i$th eigenstate of $\rho_{0}$. Similarly with the traditional
form of QFI~\cite{JLiu1,JLiu2,GRJin,JLiu3}, $I_{\theta}$ is also
determined by the support of $\rho_{0}$. For a purely initial state,
$I_{\theta}=8\langle\Delta^{2}\mathcal{H}\rangle$.

For a general exponential form state $\rho_{\theta}=\exp(G_{\theta})$,
$I_{\theta}$ is actually a correlation function, namely,
\begin{equation}
I_{\theta}=\langle\Gamma_{+}(G,\theta)\Gamma_{-}(G,\theta)\rangle,
\end{equation}
where
\begin{equation}
\Gamma_{\pm}(G,\theta):=\int_{0}^{1}\!\! e^{\pm\frac{1}{2}sG^{\times}}\!\partial_{\theta}G\: ds.
\end{equation}
When $e^{\frac{1}{2}sG^{\times}}\!\partial_{\theta}G$ is a real operator,
$\Gamma_{+}=\Gamma_{-}$. $I_{\theta}$ then reduces to $\langle\Gamma_{\pm}^{2}\rangle$.
Similarly with the SLD operator, above integrating form of $\Gamma_{\pm}$
can also be rewritten into an expanded form
\begin{equation}
\Gamma_{\pm}(G,\theta)=\sum_{n=0}^{\infty}\frac{\left(\pm1/2\right)^{n}}
{(n+1)!}\left(G^{\times}\right)^{n}\partial_{\theta}G.
\end{equation}
For a thermal state expressed in Eq.~(\ref{eq:thermal}), i.e.,~$G=-\beta H-\ln Z$,
where $\mathcal{V}$ is a conserved quantity, $\Gamma_{\pm}$ can
be expressed by
\begin{equation}
\Gamma_{\pm}=f_{\mathrm{I}}\mathcal{V}+\left(f_{\mathrm{I}}\Omega^{2}-\beta\right)
\!\partial_{\theta}H\mp2\left(\partial_{\beta}f_{\mathrm{I}}\right)
H^{\times}\partial_{\theta}H.
\end{equation}
where $f_{\mathrm{I}}$ is defined as $f_{\mathrm{I}}=2\Omega^{-3}[\frac{\beta}{2}\Omega-\sinh(\frac{\beta}{2}\Omega)].$
Similarly with $\mathcal{H}$ in unitary parametrization process,
$\Gamma_{\pm}$ is also the linear combination of $\mathcal{V}$,
the partial derive $\partial_{\theta}H$ and its commutation between
the Hamiltonian $H$. Thus, $I_{\theta}$ for thermal states of all
systems discussed previously can be calculated analytically.

\section{Conclusion\label{sec:Conclusion}}

In summary, we discuss a general scenario in which the Hermintian
operator $\mathcal{V}$ (defined in Eq.~(\ref{eq:V_definition}))
is a conserved quantity. For the unitary parametrization processes,
we provide analytical expression of the characteristic operator $\mathcal{H}$,
which is totally determined by the Hamiltonian, the commutation between
the Hamiltonian and its partial derivative, and the conserved quantity
$\mathcal{V}$. With the expression of $\mathcal{H}$, we further
give the expression of the QFI, calculate its maximum value and the
corresponding optimal initial states. For the parametrized thermal
states in this scenario, the SLD is the linear combination of $\mathcal{V}$
and $\partial_{\theta}H$.

The scenario in this paper includes many specific physical systems.
As the application, we mainly focus on two categories: $\mathfrak{su}(2)$
category and canonical category. In the $\mathfrak{su}(2)$ category,
we detailedly discuss the QFI in the ferromagnetic two-spin system,
the anisotropic two-spin XY model and a spin-one model. The characteristic
operator $\mathcal{H}$ in these systems can basically be expressed
via $\mathfrak{su}(2)$ generators. With the expressions of QFI, we
locate the optimal initial states in these systems to access the maximum
QFI. Meanwhile, the QFI for the parametrized thermal states of two-spin
systems are also discussed. In the canonical category, we provide
the QFI for a cavity optomechanical system. Increasing the photon
number in the cavity or tuning down the frequency of the movable mirror 
will enhance the QFI.

At the end of this paper, an alternative form of QFI is discussed
in the scenario. Its formula for unitary parametrization processes
is analytically given. For a general parametrized exponential state,
we also provide the expression of this alternative QFI, which is a
correlation function of $\Gamma_{+}$ and $\Gamma_{-}$. In the scenario
where $\mathcal{V}$ is a conserved quantity, $\Gamma_{\pm}$ is actually
governed by the Hamiltonian, the parameter under estimation and
the conserved quantity. We hope this paper could prompt more and more
researchers to study the connection between the conserved quantities and
the quantum Fisher information, and search for various ways to enhance the parameter precision via conserved quantities.

\begin{acknowledgments}
This work was supported by the NFRPC through Grant No. 2012CB921602
and the NSFC through Grants No. 11475146. \end{acknowledgments}

\end{document}